\documentclass[fleqn,usenatbib]{mnras}

\usepackage{newtxtext,newtxmath}
\usepackage{natbib}

\usepackage[T1]{fontenc}
\usepackage{ae,aecompl}

\usepackage{graphicx}	
\usepackage{amsmath}	
\usepackage{hyperref} 
\usepackage[capitalise]{cleveref} 

\title[DMT to expansion nebulae]{Distance mapping applied to four well-known planetary nebulae and a nova shell}

\author[G\'omez-Gordillo S. et al.]{Sebastian G\'omez-Gordillo$^{1}$,
Stavros Akras$^{1,2,3}$\thanks{E-mail: stavrosakras@gmail.com},
Denise R. Gon\c{c}alves$^{1}$\thanks{E-mail: denise@astro.ufrj.br},
\newauthor Wolfgang Steffen$^{4}$
\\
$^{1}$Observat\'orio do Valongo, Universidade Federal do Rio de Janeiro, Ladeira Pedro Antonio 43, Rio de Janeiro 20080-090, Brazil\\
$^{2}$Observat\'orio Nacional/MCTI, Rua Gen. Jos\'e Cristino, 77, Rio de Janeiro 20921-400, Brazil\\
$^{3}$Instituto de Matem\'{a}tica, Estat\'{i}stica e F\'{i}sica, Universidade Federal do Rio Grande, Rio Grande 96203-900, Brazil\\
$^{4}$Instituto de Astronom\'{i}a, Universidad Nacional Aut\'{o}noma de M\'{e}xico, Ensenada 22800, Baja California, Mexico\\
}

\date{Accepted 2020 January 6; Revised: 2020 January 5; Received 2019 May 30}

\pubyear{2020}

\begin{document}
\label{firstpage}
\pagerange{\pageref{firstpage}--\pageref{lastpage}}
\maketitle

\begin{abstract}
Accurate distance estimates of astrophysical objects such as planetary nebulae (PNe), and nova and supernova remnants, among others, allow us to constrain their physical characteristics, such as size, mass, luminosity, and age. An innovative technique based on the expansion parallax method, the so-called distance mapping technique (DMT), provides distance maps of expanding nebulae as well as an estimation of their distances. The DMT combines the tangential velocity vectors obtained from 3D morpho-kinematic models and the observed proper motion vectors to estimate the distance. We applied the DMT to four PNe (NGC~6702, NGC~6543, NGC~6302, and BD+30~3639) and one nova remnant (GK~Persei) and derived new distances in good agreement with previous studies. New simple morpho-kinematic \textsc{shape} models were generated for NGC~6543, NGC~6302, and NGC~6702, whereas for BD+30~3639 and GK~Persei published models were used. We demonstrate that the DMT is a useful tool to obtain distance values of PNe, in addition to revealing kinematically peculiar regions within the nebulae. Distances are also derived from the trigonometric {\it Gaia} parallaxes. The effect of the non-negligible parallax offset in the second {\it Gaia} data release is also discussed.
\end{abstract}




\section{Introduction}

Planetary nebulae (PNe) represent the final stage in the evolution of low- to intermediate-mass stars and they are important tools for the study of stellar evolution and gas dynamics \citep{book1}, as well as tracers of the chemical abundance in nearby galaxies \citep{magrini2012,denise2019}. PN studies rely on the knowledge of their physical properties such as the formation rate, Galactic distribution, total and ionized nebular masses, sizes, ages, luminosities and evolutionary states of their central stars (CSs), among others \citep{book2, book3, book1}. The knowledge of such physical properties is strongly dependent on the PN distances, which are still poorly determined. Traditional methods to estimate the distance to stars rely on stellar properties that are not easily applicable to PNe.

The methods used to determine PN distances are divided into two main groups, the so-called statistical methods and individual methods. The former use PN samples considering certain assumptions about the nebular structure and properties, while the latter are independent and provide a direct distance calculation to individual PN. The difference between the statistical and the individual distances can be a factor of 2 or even higher \citep{Navarro2005}. \citet{FrewParker2016} presented a new catalogue of statistical distance for 1100 Galactic PNe using an updated version of the H$\alpha$ surface brightness-radius relation \citep{2008PhDT.......109F} as well as a review of the previous statistical and individual methods. 

Accurate distance estimates are available only for few nearby PNe for which the trigonometric parallax method can be applied. The situation has changed, especially after the publication of {\it Gaia}  parallaxes, for a large number of CS PNe \citep{gaia2018}. These parallaxes will allow us to improve the calibrations of statistical distances. However, better measurements with lower fractional parallax errors are still needed \citep{Stanghellini2017,Kimeswenger2018}.

Among the individual methods \citep{book1} is the expansion parallax technique, in which the tangential velocity component ($V_{\bot}$ in~km~s$^{-1}$) and the angular expansion rate (or local internal proper motion of the gas,  $\dot{\theta}$ in~mas~yr$^{-1}$) of the nebula are used to calculate the distance (\textit{D}, in~kpc), by assuming a spherically symmetric PN, following \citep{Terzian1997},
\begin{equation} \label{ep}
D~[\text{kpc}] = 211\frac{V_{\bot}~[\text{km~s}^{-1}]}{\dot{\theta}~[\text{mas~yr}^{-1}]} 
\end{equation}

In general, PNe display a great diversity of shapes and morphologies as the result of the complex and not yet fully understood radiative and hydrodynamical mass-loss evolution of their progenitors \citep{Balick2002}. According to \citet{schonberner2005evolution} the global structure of spheroidal PNe can be understood in terms of three nebular components that are kinematically and morphologically well distinguishable: the rim, the shell and the halo, all surrounding the CS.

\citet{reed1999hubble} showed that the physical radial expansion velocity of a nebula, obtained spectroscopically, can be converted to tangential velocity by applying a morpho-kinematic model. This approach is appropriate since several regions in PNe may not meet the requirement for \cref{ep} of having equal radial and tangential expansion velocity due to the lack of spherical symmetry \citep{li2002angular}. The constructive morpho-kinematic modelling involves only structural (morphological) and velocity (kinematic) information from the nebula, allowing to reconstruct it in a simple and broad way or in a complex and more detailed manner. This is possible due to the availability of high spatial resolution imaging and high-dispersion spectroscopy, fundamental tools to constrain the kinematics and morphology of the nebular models.

Pursuing a distance calculation by means of a morpho-kinematic analysis of the expanding nebulae, a novel variation of the expansion parallax technique, the distance mapping technique (DMT), was introduced and first applied to BD+30~3639 by \citet{akras2012}. The novelty of the DMT lies in applying \cref{ep} multiple times, in various regions within the nebula, to create a distance map. This allows to better constrain 3D morpho-kinematic models of expanding nebulae and determine the nebular distance, by using observed proper motion vectors and modelled velocity fields. The average distance of the non-zero values derived from the distance map corresponds to the distance of the nebula. The error map and the error of the average distance are derived using the error propagation methodology. 

The DMT is a new addition to the set of 3D distance determination techniques. 3D photoionization modelling has also been used to derive the distance of PNe using as constraints a number of observable parameters like geometry, size, extinction map,  density map, emission-line fluxes, stellar temperature, and luminosity \citep{monteiro2004,monteiro2005,monteiro2006,monteiro2011,akras2016}.

In this paper, we apply a new version of DMT to a number of PNe not only to estimate the distances but also to demonstrate that it serves to get new insights into the kinematics and even the morphology of an expanding nebula. In \cref{DMTt}, we present some upgrades of the code along with the advantages that this technique has when studying the kinematics of an expanding nebula. The kinematics and morphological properties of the PNe involved in this paper are introduced in \cref{KM}. In \cref{M}, we present our morpho-kinematic modelling approach for the objects with available proper motions using the 3D morpho-kinematic code \textsc{shape} \citep{steffen2006,steffen2011shape}. The results from the DMT application and their comparison with those in the literature are discussed in~\cref{APP}. Finally, we present our results and conclusions in \cref{F}.

\section{The Distance Mapping Technique}
\label{DMTt}

Since the debut of the DMT \citep{akras2012}, some upgrades have been made and two of them are worth mentioning. The first upgrade deals with the procedure to determine and exclude the outliers in the distance estimation. The outliers are found by means of the Z-score method \citep{iglewicz1993detect} where a data point is described in terms of its relationship with the median absolute deviation and the median value of the sample. By adding this function more reliable distances are expected. The second upgrade in the DMT is related with the resolution of the distance map. Due to the changes in the distribution of the distance and error with the resolution of the maps, the new version provides those distance and error maps with a resolution for which the intrinsic error of the DMT is minimum. The resultant resolution of the maps is derived by minimizing the ratio between the distance dispersion and the mean value.

The most important input parameters in the DMT are the sets of (i) modelled tangential velocity vectors and (ii) observed proper motion vectors. The first set is derived from a morpho-kinematic model of the nebula -- the 3D morpho-kinematic code \textsc{shape} is an example of a software used to create those models. The second set corresponds to the observational proper motions derived from images obtained at two (or more) different epochs \citep{hajian2006distances}. The primary outputs of the DMT are the distance and error maps.

From a perfect match between the modelled tangential velocity field and the observed proper motions, homogeneous distance and error maps are expected. Deviations, however, will arise due to observational errors as well as deviations of the morpho-kinematic model from the actual nebula. In the case that several adjacent cells in the distance map display a similar systematic deviation, they form a region called systematic distance deviation region (SDR). These SDRs likely correspond to localized divergence between the 3D modelled velocity and the observed proper motion fields. The nature of these divergences may be related to additional physical phenomena in the nebula, which are not taken into account in the morpho-kinematic models, or with problematic observations. If all observational reasons for the deviation have been excluded, the SDR may be used as a constraint for model improvements. 

The final error of the DMT distance is the combination of three components: (i) the intrinsic error of the technique based on the number of proper motion measurements and tangential velocities in each cell; (ii) the observational error of the proper motions; and finally (iii) the error of the correction factor. Besides the DMT average distance of each source, we also provide (when suitable) the 1$\sigma$ distance range based on the dispersion of the distance distribution in the map.

The intrinsic error determined for each cell is the statistical error of the mean values of the modelled tangential velocity and the observed proper motion. In general, the former is higher than the latter due to the small number of observed proper motion measurements that lie in each cell in comparison with the large number of modelled tangential velocities. The dominant contributor to the distance error is the observational error of the proper motion measurements. In the absence of observational errors for the proper motions, the distance error represents only the intrinsic uncertainty of the DMT. Ideally, a better error estimation is expected when a set of observed errors for each proper motion measurement is available. Only for GK~Persei such a set was available in the literature, while for the remaining objects an average observational error value for the whole set of measured proper motions is given.


Unfortunately, there is no easy way to quantify the fitting quality of a 3D morpho-kinematic model to the observed data; therefore, this is an extra uncertainty in the distance that is not taken into account. Nevertheless, through a simulated grid of simple \textsc{shape} models, we explore how the resultant distances vary relative to a reference value as a function of the number of tangential velocities and proper motions as well as the position and inclination angles of the modelled nebulae. The deviation of the DMT distances relative to the reference value was found to be (i) up to 20\% for models with less than 100 proper motion measurements, which though can decrease until 5\% for the models with more than 100 proper motions; (ii) up to 5\% for the models with a difference in position angle (PA) from the observations less than 15~º; (iii) 2-3\% for the models with inclination angle difference less than 10~º; and finally (iv) the error of the resultant distances decreases with the resolution of the maps (i.e. number of cells in the map).  

Overall, small deviations between the simulated and true nebula yield an intrinsic distance difference from the true distance up to 5-7\%. This difference may be even higher if the number of available observed proper motions is small, such as the case of NGC~6720 (see later).

\citet{mellema2004expansion} has shown that the spectroscopically measured Doppler velocity (radial) and angular motions of an expanding PN may not correspond to the same physical component in the nebula. In particular, the former motion refers to the expansion of the ionized gas behind the ionization front (which is called matter velocity), whereas the angular motion corresponds to the expansion of the ionization front itself (which is called pattern velocity). A difference up to 30\% between the pattern and matter velocities has been reported, entailing an underestimation of the expansion parallax distance. \citet{schonberner2005evolution} have also reached the same conclusion using 1D hydrodynamical modelling of PNe. Given that the DMT relies on the expansion parallax technique, the resultant distance must be corrected, by applying a correction factor.

\section{KINEMATIC AND MORPHOLOGICAL DESCRIPTION}
\label{KM}

The morpho-kinematic study of PNe with high spatial resolution imagery and high-dispersion spectroscopic data allows us to better understand and gain insights into the formation, interaction and evolution of various nebular components (e.g. shell, rim, bipolar outflows, knots, torus, etc.) Next, we give a short description of the morpho-kinematic characteristics of our sample of PNe.

\subsection{NGC~6543}
\label{sec:maths} 

A morphological description of NGC~6543 or better known as the Cat's Eye Nebula was provided by \citet{reed1999hubble} with a detailed description of the inner and outer features. An inner elongated bubble (a prolate ellipsoid), called the inner ellipse, is located at the centre of the nebula and it is identified by the hot gas that fills it. Its temperature, derived from X-ray spectral fitting, is of the order of $1,700,000\ K$ \citep{chu2001}. It displays a non-uniform expanding behavior, with the outermost tips having the lowest velocities. This inner ellipse turns out to be embedded in a larger bipolar structure, a pair of large spherical bubbles that are conjoined, where the binding region of the bubbles forms the waist of the nebula. The waist is observed as a second larger ellipse lying perpendicular to the inner ellipse. A pair of polar caps and polar filaments are also present at the tips of the bubbles, being very bright in low-ionization lines \citep{goncalves2001,akrasdenise2016} and molecular hydrogen lines (Akras et al. (2019), submitted). The conjoined bubbles are older and expand with a lower velocity compared to the inner ellipse \citep{reed1999hubble}.

\subsection{NGC~6720}

The Ring Nebula has a triaxial pole-on shape oriented towards us and it is composed of a thick elliptical equatorial ring (called the main ring), plus a thin and faint polar region \citep{odell2007}. The main ring is an elliptical structure of ionization bounded gas with a strong density concentration in its equatorial plane. In addition, it is surrounded by a glow of [\ion{O}{iii}]  emission and a low-ionization halo. In contrast to the bright nebular emission, several dark knots are seen, pointing towards the CS \citep{odell2013}. 

A homologous expansion has shown an excellent agreement with the measurements of the radial and tangential velocities of [\ion{N}{ii}] emission-line features, the dark knots and  the ionized gas \citep{odell2007,odell2013,odell2009}. In such scenario, ions of higher ionization expand more slowly being such type of motions commonly found in PNe \citep{odell2007,akras2012,akras2015}.

\subsection{NGC~6302}

The complex morphology of NGC~6302 or the Butterfly Nebula is portrayed with several emission lines in the WFC3 HST image of \cite{szyszka2011}. Two prominent lobes emerge from the torus in the eastern and western directions. A homologous expansion velocity law has been confirmed in the lobes by \citet{meaburn05} using morpho-kinematic modelling of spatially resolved optical line profiles plus the proper motion measurements of 15 knots in the north-western lobe. \citet{meaburn05} reported outflow velocities $\geq 600$~km~s$^{-1}$ for the north-western lobe. A homologous velocity law was also reported by \citet{peretto2007} in the CO emission close to the expanding torus. The inner nebular region presents evidence of a recent additional acceleration, more pronounced to the southern region probably due to the overpressure after the beginning of the photoionization \citep{szyszka2011}.

\subsection{BD+30~3639}

\citet{bryce1999} established the morphological structure of BD+30~3639 by using spatially resolved, high spectral resolution observations. These authors found that this PN is composed of a main nebular shell of low-ionization emission, almost axially symmetric, prolate and with an open ended shape, together with a closed shell structure of smaller size and of high-ionization emission. Following these authors, the [\ion{O}{iii}] expansion velocity ($\sim$ 35.5~km~s$^{-1}$) is higher than the [\ion{N}{ii}] one ($\sim$ 28~km~s$^{-1}$)  indicating a non-homologous expansion law for BD+30~3639. \citet{akras2012} argued for the presence of collimated outflows or winds in BD+30~3639. This is supported by the CO bullets with expansion velocities of $\sim$50~km~s$^{-1}$ along the polar direction \citep{bachiller2000}. A decreasing expansion velocity function for internal higher ionization ions and an increasing expansion velocity function for external low-ionization ions were found by \citet{sabbadin2006}. Such velocity law generates a `V-shaped' velocity profile across the nebula. In this scenario, the expansion velocity is high close to the nebular centre but decreases quickly with the nebular radii and finally increases again to the external regions of the nebula. \citet{gesicki2003} discussed 10 Galactic PNe with `V-shaped' velocity profiles based on the emission lines of \ion{H}{$\alpha$}, [\ion{N}{ii}] and [\ion{O}{iii}]. 

\subsection{GK~Persei}

GK~Persei, also known as the Firework Nebula is a nova remnant formed by the material ejected during an outburst in a cataclysmic variable star (CV). CVs are short-period interacting binary systems that consist of a white dwarf star (WD) accreting mass from a main-sequence star, which has filled its Roche lobe. Eventually, an outburst eruption on the surface of the WD is generated by thermonuclear runaway of the accretion material \citep{warner1995}. Classic novae eject large amounts of mass, around $10^{-4}$~M$_\odot$ \citep{bildsten2004}, at velocities of the order of 500-5000~km~s$^{-1}$ \citep{bode2008}. The CV of GK~Persei consists of an evolved late-type star (K2IV) \citep{crampton1986} together with a magnetic WD \citep{bianchini1983}. The remnant of this mass ejection has a total mass of $7\times10^{-5}$~M$_\odot$ \citep{pottasch1959v} and an electron temperature greater than $2.5\times10^{4}$~K \citep{williams1981}. The shell of the nebula has been described as clumpy, asymmetric, and boxy \citep{anupama2005}, and it is interacting with the environment by suffering a deceleration at the south-west region due to shocks \citep{seaquist1989, bode2004}. More recently, by means of morpho-kinematic modelling, its remnant was described as a cylindrical shell with embedded lower velocity polar structures \citep{harvey2016}.

\section{\textsc{Shape} morpho-kinematic modelling}
\label{M}

Over the past 10-15 yr, the code \textsc{shape} \citep{steffen2006,steffen2011shape} has been used to study the morphological structure and expansion velocity field of several PNe, in 3D, with special focus on high-velocity collimated outflows \citep{,akraslopez2012,akras2012,akras2015}, bipolar nebulae \citep{clyne2014,clyne2015}, complex multipolar geometries \citep{clark2010,clark2012,chong2012,hsia2014,gomez-munos2015,rubio2015,sabin2017}, or high-velocity knots \citep{vaytet2009,derlopa2019}. Next, we discuss in details the morpho-kinematic \textsc{shape} model of each PN studied here.
	
\subsection{NGC~6543}
\label{6543m}
\begin{figure}
	\includegraphics[width=\columnwidth]{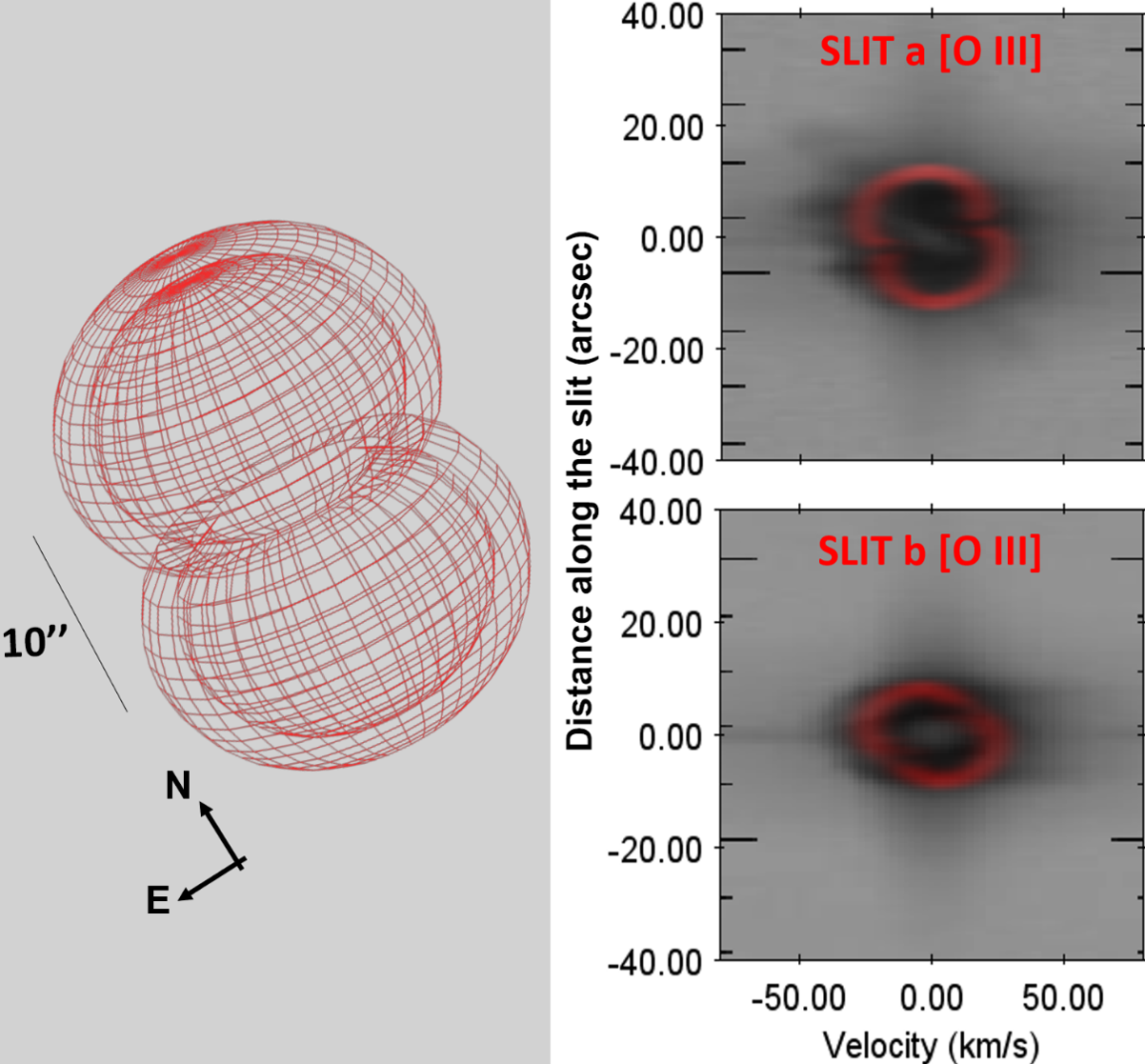}
	\caption{3D bipolar model of NGC~6543 (left). Synthetic [\ion{O}{iii}] PV diagrams in red overlaid on the observed ones \citep{lopez2012} (right).}
	\label{Fig:composed6543}
\end{figure}

\textsc{shape} was used to reconstruct the 3D structure of NGC~6543 based on the structure's description of \citet{reed1999hubble} and position-velocity (PV) diagrams obtained from the San Pedro Martir catalogue of Galactic PNe \citep{lopez2012}, by using only the slits \lq a\rq and \lq b\rq of the [\ion{N}{ii}] and [\ion{O}{iii}] emission lines. A simple geometrical approach was assumed for the 3D structure of this nebula. The two conjoined bubbles were modelled assuming a bipolar structure, whose lobes are simulated as spherical shells with a constant density (\cref{Fig:composed6543}, left-hand panel). 

The [\ion{N}{ii}] image and the PV diagrams were used to reproduce the 3D morpho-kinematic model, despite the fact that the proper motions were measured from the [\ion{O}{iii}] image. It was necessary to construct a 3D model using the [\ion{N}{ii}] PV diagrams since this emission line provides a more detailed view of the nebular components of NGC~6543 than the [\ion{O}{iii}] PV diagrams. To create the final [\ion{O}{iii}] 3D model, the [\ion{N}{ii}] 3D model was scaled to match the [\ion{O}{iii}] PV diagrams. 

For the expansion velocity field,1 we considered a single homologous velocity law, with $V = 22*(r \setminus r_o)$~km~s$^{-1}$, where $r$ is the distance from the CS and $r_o$ is the reference radius of the shell. The synthetic PV diagrams generated by our model (red) are overlaid to the observed ones (right panels, \cref{Fig:composed6543}) displaying an acceptable agreement with the observed PV diagrams, despite the simplicity of the model.

\subsection{NGC~6720}
\label{6720m}
Three scenarios have been proposed in order to explain the 3D structure of this nebula: (i) a prolate ellipsoid with open ends along the major axis, which resembles a barrel structure \citep{guerrero1997, hiriart2004,odell2007}; (ii) a bipolar nebula nearly pole-on \citep{bryce1994,kwok2008}; and finally (iii) a combination of the two aforementioned models in which the PN has an open-end barrel-shaped central region that is viewed along its axis \citep{sahai2012}.

\begin{figure}
	\includegraphics[width=\columnwidth]{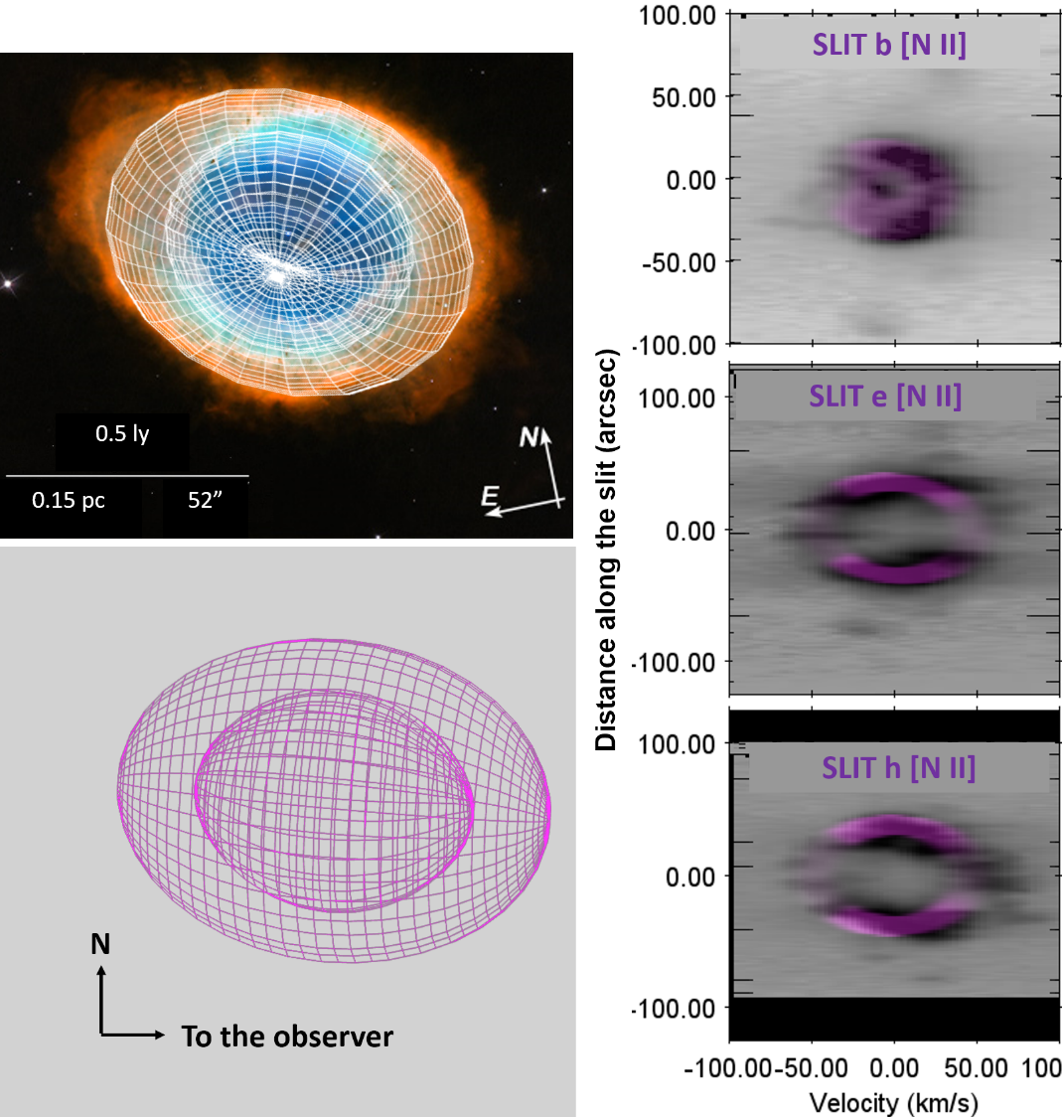}
	\caption{Our 3D model superimposed on the NGC~6720 composite image of \citet{odell2013} (top left) and the curved cone seen edge on (bottom left). Synthetic PV diagrams in purple overlaid on the observed ones \citep{lopez2012} (right).}
	\label{Fig:composed6720-2}
\end{figure}

Our morpho-kinematic model of NGC~6720 follows a much simpler approach. Some features are taken from models (i) and (iii) resulting in a prolate ellipsoid shell seen pole-on (\cref{Fig:composed6720-2}, left-hand panel), primarily modelling the main ring. A single homologous velocity law $V = 37(r/r_\text{0})$~km~s$^{-1}$ was selected for the model based on the observational PV diagrams. The \textsc{shape}  modelled PV diagrams are presented in the right-hand panel of \cref{Fig:composed6720-2}, for three different slit positions, superimposed upon the observed PV diagrams obtained from SMP catalogue \citep{lopez2012}. Some small deviations between the modelled and observed PVs can be observed, which are attributed to the protrusion of the real structure of the nebula compared to our ellipsoidal model, as can been seen in the left-hand panel of \cref{Fig:composed6720-2}. 

\subsection{NGC~6302}
\label{m6302}
The first attempt to model the 3D structure of NGC~6302 was carried out by \citet{meaburn05} assuming a homologous velocity law and rotationally symmetric lobes along with two deformed spheroidal surfaces that aimed at modelling the knotty structures. Although this model successfully reproduced the observed [\ion{N}{ii}] PV diagrams of NGC 6302, the current \textsc{shape} version (v. 5.0) is not compatible with the models developed  using  older versions. For this reason, we had to construct a new shape model of NGC 6302.

\begin{figure}
	\includegraphics[width=\columnwidth]{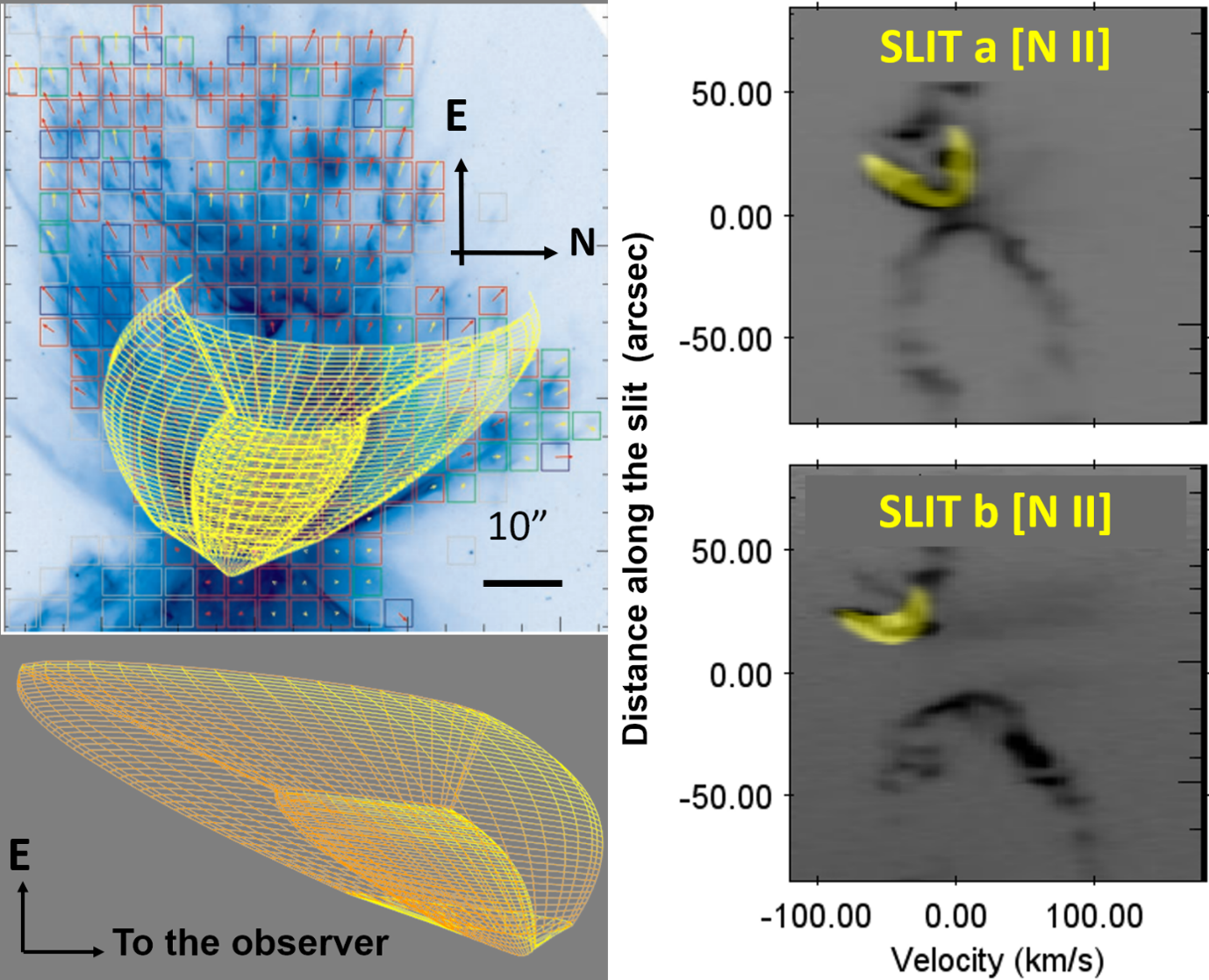}
	\caption{Our 3D model of NGC~6302 properly oriented, overlay the proper motion velocity field of the eastern lobe \citep{szyszka2011} (top left), and the conical shell seen edge on (bottom left). Observed and synthetic (yellow) PV diagrams of the slits \lq a\rq and \lq b\rq \citep{meaburn05} (right).}
	\label{Fig:composed6302}
\end{figure}

We decided to model again the eastern lobe of the nebula considering an extended 3D conical shell (\cref{Fig:composed6302}, left-hand panel), because the proper motion velocity field is available only for this lobe \citep{szyszka2011}. Our 3D morpho-kinematic model presents a single velocity law, $V = 28(r/r_0)$~km~s$^{-1}$ and approximately reproduces the [\ion{N}{ii}] PV diagrams of \citet{meaburn05} for slits `a' and `b' (\cref{Fig:composed6302}, right-hand panel). The synthetic PV diagrams adequately reproduced the main features, although without taking into account the clumps and voids, which is not necessary for the DMT.

\section{DMT application}
\label{APP} 

With the 3D morpho-kinematic \textsc{shape} models constructed, the tangential velocity field for each PN is obtained. The observed proper motion measurements and the tangential velocities are used as input to the DMT to derive the distance maps for each nebula. In the following sections, we discuss the DMT results and the comparison with the distances available in the literature.

At this point, it has to be mentioned that four objects in our sample have available \textit{Gaia} parallaxes in the second data release \citep{gaia2018}. Besides their DMT distances, we also derived new distances from the inverse of their {\it Gaia} parallaxes which is feasible because of the low fractional errors \citep[$<$20 per cent, as recommended by][]{gaia2018,Luri2018}. However, parallaxes in the {\it Gaia} DR2 suffer from some systematic errors. There is no simple way to quantify these errors, which depend on the position of the sources in the sky, their magnitudes, and their colours, but they are of the order of 0.1~mas. For this work, we make use only of the uncertainties provided in the {\it Gaia} DR2, as recommended by \cite{Luri2018}. This means that the errors of the distances derived from the {\it Gaia} parallaxes provided in Tables~1-5 are likely larger, as also pointed out by \cite{Kimeswenger2018}.

Although {\it Gaia} should provide absolute parallaxes, a non-negligible shift in parallaxes has been reported in the DR2, which has to be taken into consideration. As reported by the {\it Gaia} team, this parallax offset is in the range from 10 to 100 mas, depending on the magnitude, colour, and position of the sources in the sky \citep{arenou2018,lindegren2018}. Several independent studies have been carried out to trace and determine this parallax zero point using different types of sources. For instance, \cite{lindegren2018} determined a parallax zero point of 0.029~mas using a sample of quasars \citep{lindegren2018}. \cite{reiss2018} and \cite{Groenewegen2018} reported independently parallax zero-points of 0.046$\pm$0.011 and 0.048$\pm$0.018 mas, respectively, from samples of Cepheids. By the study of a sample of RR-Lyrae, the parallax zero point was found to be 0.057$\pm$0.0034~mas \cite{Muraveva2018} comparable with the values derived from Cepheids. \cite{Stassun2018} reported a parallax offset of 0.082$\pm$0.033~mas using a sample of eclipsing binaries. In 2019, many more studies dealt with the {\it Gaia} parallax zero point and the value varies between 0.030 and 0.1~mas (0.031$\pm$0.11mas, \cite{graczyk2018}; 0.054$\pm$0.006mas, \cite{Schonrich2019}; 0.0523$\pm$0.002 mas, \cite{Leung2019}; 0.075$\pm$0.029mas, \cite{Xu2019}; 0.0517$\pm$0.0008mas, \cite{Khan2019}; 0.03838$^{+0.01354}_{-0.01383}$~mas, \cite{hall2019}; and 0.0528$\pm$0.0024 mas, \cite{zinn2019}).

It is therefore undoubtable that there is a non-zero parallax offset in the {\it Gaia} DR2 between 0.2 and 0.8~mas. There is a significant difference in the parallax zero-points derived from fainter and bluer quasars (0.029 mas, \cite{lindegren2018}) and from brighter and redder sources like Cepheids (0.046~mas, \cite{reiss2018}; 0.048~mas, \cite{Groenewegen2018}), RR Lyrae (0.057~mas, \cite{Muraveva2018}), or red giants and red clump stars (0.0528~mas, \cite{zinn2019}; 0.0517~mas, \cite{Khan2019}; and 0.03838~mas, \cite{hall2019}). 

Even though PN CSs are bluer sources and the zero point derived from quasars might be better suited, the CSs of PNe are significantly brighter (G$\leq$12) than quasars. Because of the unique characteristics of PN CSs (magnitudes/colours), an independent work is necessary to properly determine the parallax zero point in the {\it Gaia} DR2 for hot and bright stars. 

For this work, we have adopted the zero point derived from quasars (0.029 mas, zero point correction A) and the weighted average (0.051 with standard deviation of 0.025, zero point correction B) calculated from all the aforementioned works.

\subsection{NGC~6543}
 
\begin{figure}
	\includegraphics[scale=0.42]{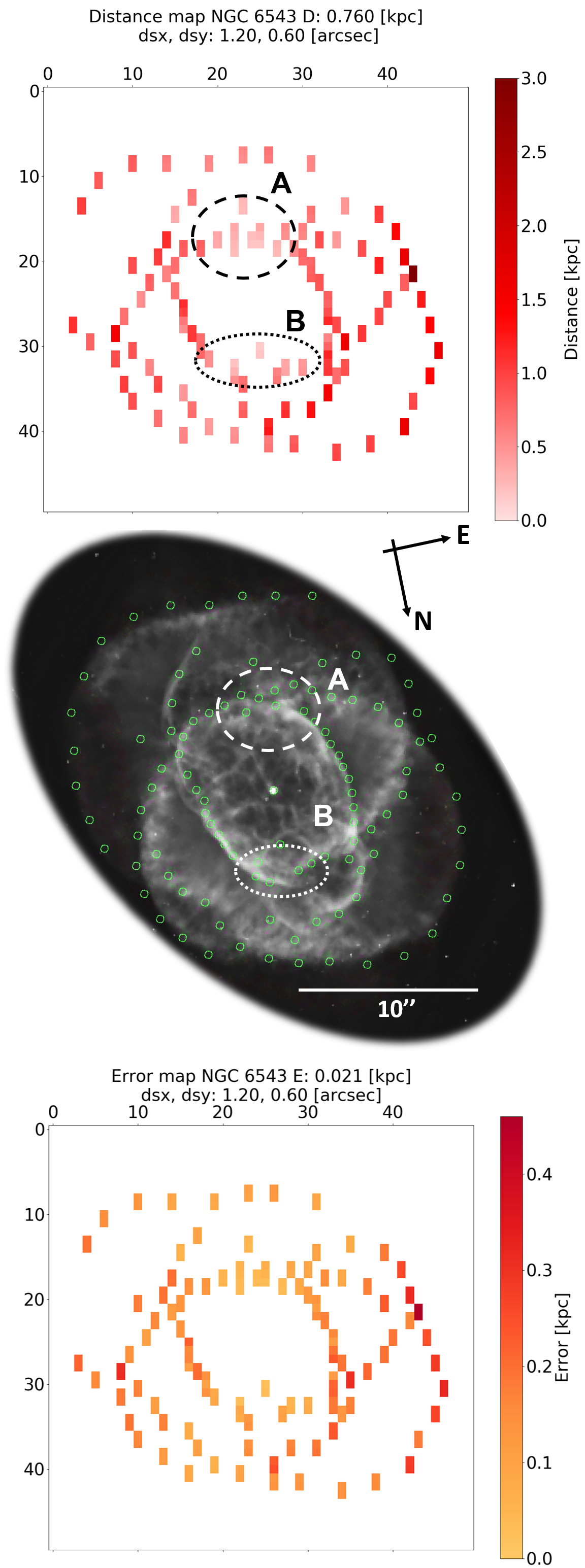}
	\caption{NGC~6543 distance map (top panel) on which SDR A and B are marked in black. The mean distance value (\textit{D}) = 0.76~kpc is given in the title. The [\ion{O}{iii}] HST image showing the location of the 107 measured proper motions (green circles) with the SDRs in white (middle panel). The error map shows higher error for the cells with higher distances (lower panel).}
	\label{Fig:dmt6543}
\end{figure}

For this PN, 107 proper motion measurements in the [\ion{O} {iii}] emission line with an observational error of 20\% for each proper motion \citep{reed1999hubble} and the modelled velocity field obtained in Section \ref{6543m} were used. The resultant distance map displays dispersed and aggregated deviations from the mean distance (\cref{Fig:dmt6543}, top panel). Two systematic distance deviation regions or SDRs, labeled as A and B, are identified (\cref{Fig:dmt6543}, bottom panel). The discrepancy between $V_{\bot}$ (the tangential modelled velocity component from a homogeneous field) and $\dot \theta $ (observed velocity component) results in the distance deviations. In the case that the distance of an SDR is above the mean distance of the nebula, then the measured proper motions are lower than the expected values from the modelled tangential velocity vectors. On the contrary, for an SDR distance below the mean value, the measured proper motions are higher than the modelled tangential velocity vectors. 

We find that both SDRs A and B are in the regions where the inner ellipse and the bipolar waist are crossed. It is likely that the measured proper motions in these regions are associated with both the bubbles and the inner ellipse. Hence, the modelled velocity field (only the bubbles are modelled) diverges from the associated proper motions in these regions. Extreme low and high values of the distance map are excluded from the estimation of the mean value, as outliers, to ensure that deviations coming from the simplicity of our model will not have significant impact on our final estimation of the distance.

For this work, the correction factor of 1.56 $\pm$ 0.15 \citep{mellema2004expansion,schonberner2005evolution} is adopted. This yields a distance to NGC~6543 of $1.19\pm0.15$~kpc (see \cref{DMTt}). It is found to be very close to the value derived from \cite{FrewParker2016}, but it does not agree with the more recent values of 1.86 and 1.612~kpc reported by \cite{schonberner2018} and \cite{Schonberner2019}. Besides the mean DMT distance value, we also yield the most probable 1$\sigma$ distance range between 0.9 and 1.48~kpc, using the dispersion of the distance map ($\sigma_D = 0.29$~kpc). This distance range encloses 10 out of the 20 distances from the literature.

NGC~6543 has also available parallax measurement in the {\it Gaia} DR2, with low fractional error. Its distance is estimated, from the inverse parallax, to be equal to 1.625~kpc. This trigonometric distance value is very close to the distances reported by \cite{mellema2004expansion},\citet{reed1999hubble}, and \cite{Schonberner2019}, but substantially different from our DMT distance of 1.19~kpc or the one reported by \cite{FrewParker2016} of 1.15~kpc or by \cite{Schonberner2019} of 1.86~kpc. If we take into account the reported non-zero parallax offsets in the {\it Gaia} DR2 (see \cref{DMTt}), the new corrected {\it Gaia} distances become smaller than and equal to 1.501 and 1.552~kpc, depending on the parallax offset.

The distance of 1.86~kpc reported by \cite{schonberner2018} is the highest among all the distance estimations over the last 20 yr (see \cref{table:d6543}) and obtained from the average of the [\ion{N} {ii}] and [\ion{O} {iii}] emission lines, 1.98$\pm$0.20 and 1.73$\pm$0.18~kpc\footnote{The names of the emission lines are in reverse (see \citet{schonberner2014} and \citet{schonberner2018})}, respectively. The former is found to be very high and inconsistent with the previously published values, while the latter agrees within its uncertainties. For these distance estimations, the corrected factor, F, was equal to 1.4 and 1.6 for the two emission lines, while for the new distance equal to 1.612~kpc reported by \cite{Schonberner2019}, the correction factor is 1.3.

It should be noted that the complexity of NGC~6543 (see section 3.1) does not allow to describe it by 1D hydrodynamic models and obtain reliable correction factors \citep[e.g.][]{schonberner2018}. Moreover, from the upper panel of \cref{Fig:dmt6543}, one can see that the individual distance estimations for NGC~6543 can vary significantly from 0.5 up to 1.5~kpc. This signifies that the expansion velocity of the gas is so complicated that the calculation of the pattern and matter velocities from one region in the nebula may not be adequate and result in incorrect distance, despite the high quality of the observations. 

Based on this analysis, it is evident how difficult it is to obtain reliable distances to PNe, given the uncertainty of the parameters involved in these estimations. Taking into account all the distance estimations available for NGC~6543 (\cref{table:d6543}), we argue that the probable value should be within the 1.3 and 1.6~kpc range.

\begin{table}
	\centering
	\caption{Cat's Eye distance determinations.}
	\begin{tabular}{ c c }
		\hline
		Distance (kpc) & Reference \\ 
		\hline
		1 & \citet{cahn1971} \\ 
		1.17 &  \citet{castor1981} \\  
		4.14 &  \citet{phillips1984} \\
		0.7 & \citet{maciel1984} \\
		0.64 & \citet{cahn1984} \\
		0.89 & \cite{kaler1985} \\
		1.39 & \citet{bianchi1986} \\
		1.44 & \citet{perinotto1989} \\
		1.8 & \cite{malkov1997} \\
		1.001 $\pm$ 0.269$^{\textit{a}}$ & \citet{reed1999hubble} \\
		0.9 - 1.3 & \citet{cazetta2001} \\
		1.55 $\pm$ 0.44 & \citet{mellema2004expansion} \\
		1.14 & \citet{phillips2005} \\
		1.607 $\pm$ 0.3213 & \citet{Stanghellini2010}\\
		1.15 $\pm$ 0.32 & \citet{FrewParker2016} \\
		1.86 $\pm$ 0.15 & \citet{schonberner2018} \\
	    1.536$^{\textit{b}}$ & \citet{Bailer2018} \\	
	    1.612 $\pm$ 0.25 & \citet{Schonberner2019}\\
		1.625 $\pm$ 0.187$^{\textit{c}}$ & {This work - no zero-point correction}\\
		1.552 $\pm$ 0.179$^{\textit{d}}$ & {This work - zero-point correction A} \\
		1.501 $\pm$ 0.173$^{\textit{e}}$ & {This work - zero-point correction B} \\
		\hline		
		1.19 $\pm$ 0.15 & This work - DMT\\
		0.9 - 1.48 & Most probable distance range - DMT\\	
		\hline   
	\end{tabular}
	\label{table:d6543}
	\begin{flushleft}
	$^{\textit{a}}$ This distance value is not corrected for the difference between the matter and pattern velocities (correction factor, \textit{F}). If \textit{F} is applied, the distance will be 1.56~kpc.\\
	$^{\textit{b}}$This value is obtained from the catalogue of geometrical distances following a statistical approach. The lower and upper bounds are 1.382 and 1.727~kpc, respectively, on the confidence interval of the estimated distance.\\
	$^{\textit{c}}$ This value is derived from the inverse {\it Gaia} parallax of 0.6152 $\pm$ 0.0709~mas. The fractional error is as low as 0.12 allowing us to derive a reliable distance using the inverse parallax.\\
	$^{\textit{d}}$ This value is derived taking into account a parallax zero point of 0.029~mas in the {\it Gaia} parallaxes using a sample of quasars \citep{lindegren2018}.\\
	$^{\textit{e}}$ This value is derived taking into account a weighted average parallax zero point of 0.051~mas in the {\it Gaia} parallaxes (see text). \\
	
	\end{flushleft}
\end{table}

\subsection{NGC~6720}

\begin{figure}
	\includegraphics[scale=0.39]{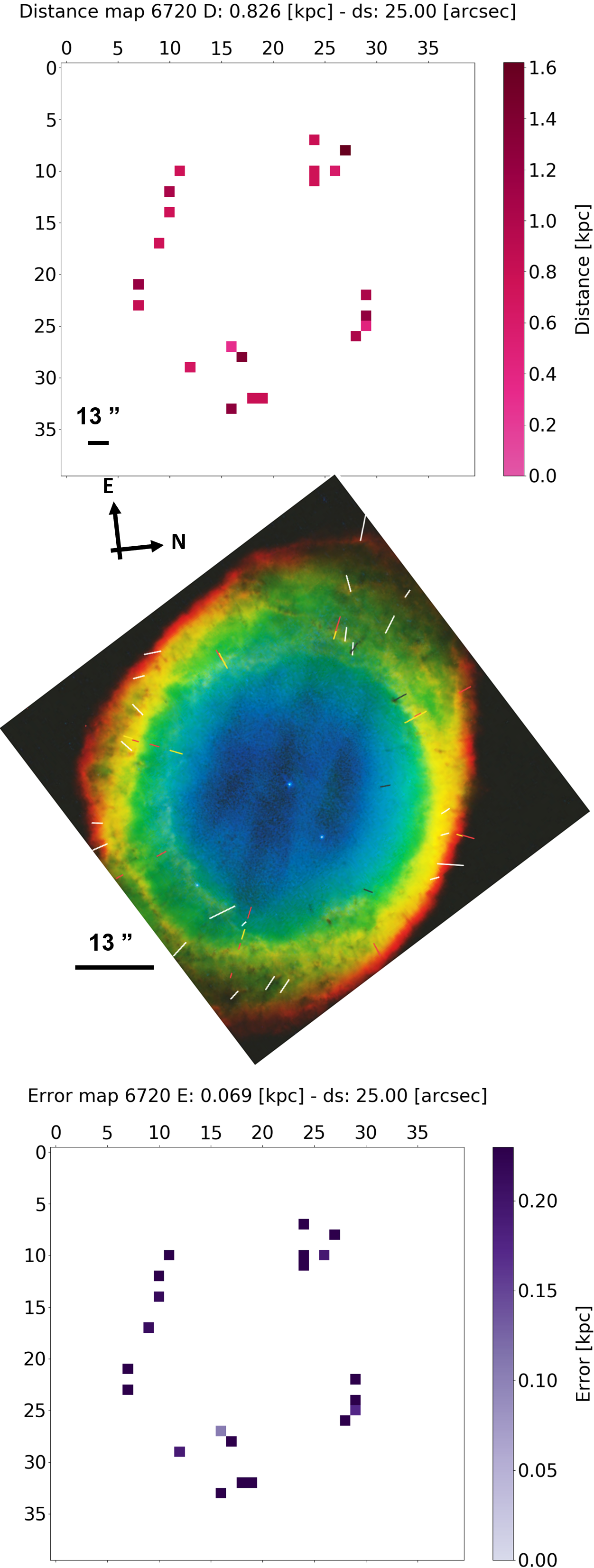}
	\caption{Distance map for NGC~6720 (top panel). Composite HST image of NGC~6720 with 22 measured proper motions \citep{odell2009} shown as the white coloured bars (middle panel). As in the error map of NGC~6543, NGC~6720 displays higher errors for larger distances (lower panel).}
	\label{Fig:dmt6720}
\end{figure}

For this PN, there are only 22 [\ion{N}{ii}] proper motion measurements with an observational error of 40\% \citep{odell2009}. Due to the low number of proper motions, the distance map (\cref{Fig:dmt6720}, top panel) is not morphologically relevant. Hence, kinematic analysis of the distance map in terms of SDR is not possible. The DMT mean distance of NGC~6720 is calculated to be 826~pc. This value must be corrected due to the systematic error of the expansion parallax technique \citep{mellema2004expansion,schonberner2005evolution}. The correction factor for NGC~6720, $F = 1.3\ \pm\ 0.26$, was derived from Figure 11 in \cite{schonberner2005evolution} and the surface temperature of its CS. The final distance estimation is $1.07\pm\ 0.3$~kpc. We find that the DMT distance is in agreement with 9 out of 17 values listed in \cref{table:d6720} (see \cref{Fig:Tplot}). We do not provide the 1$\sigma$ distance range for this nebula because the final distance error is larger than $\sigma_D$. Our value is higher than the distances obtained by \citet{harris2007} and \citet{odell2009,odell2013}, but it agrees within its uncertainties with the value reported by \citet{2008PhDT.......109F} and \citet{FrewParker2016}.

\textit{Gaia} distances (corrected or non-corrected for the parallax offset) are found to be lower than our value, being consistent with the values reported by \citet{odell2009,odell2013} and slightly out of our $1\sigma$ distance range. Nevertheless, our distance value and those from \citet{odell2009,odell2013} are also in agreement, if the high uncertainties of these estimations are taken into account. From \cref{Fig:dmt6720}, one can see that there are a few pink cells that correspond to a distance between 0.6 and 0.8~kpc as well as dark pink cells with distances between 1 and 1.4~kpc. A mean distance value of 0.826~kpc is obtained before applying the correction factor. Apparently, the small number of proper motion measurements (22, \citet{odell2009}) does not allow us to better constrain the distance of this nebula for which a discrepancy of up to 20 percent from the real value is possible (see \cref{DMTt}).
	
\begin{table}
	\centering
	\caption{NGC~6720 distance values estimated over the last 63 yr.}
	\begin{tabular}{ c c }
		\hline
		Distance (kpc) & Reference \\ 
		\hline 
		0.39 & \citet{shklovsky1956new} \\
		0.676 & \citet{O'Dell1962} \\
		0.81 & \citet{zuckerman1980} \\
		0.872 & \citet{cahn1992} \\
		0.5$\ \pm\ ^{0.21}_{0.11}$& \citet{pier1993} \\
		1.13 & \citet{zhang1995} \\
		1.11 & \citet{napiwotzki2001} \\
		0.45 & \citet{odell2002} \\
		0.7$\ \pm\ ^{0.45}_{0.2}$ & \citet{harris2007} \\
		0.9 & \citet{2008PhDT.......109F} \\
		0.74$\ \pm\ ^{0.5}_{0.2}$ &  \citet{odell2009} \\  
		$0.720\pm 30\%$ &  \citet{odell2013} \\
		0.92 $\pm$ 0.26 & \citet{FrewParker2016} \\
	    0.771$^{\textit{a}}$ & \citet{Bailer2018} \\
		0.787 $\pm$ 0.037$^{\textit{b}}$ & {This work - no zero-point correction} \\
		0.769 $\pm$ 0.036$^{\textit{c}}$ & {This work - zero-point correction A} \\
		0.757 $\pm$ 0.035$^{\textit{d}}$ & {This work - zero-point correction B} \\
	    \hline 
		1.07 $\pm$ 0.3$^{\textit{e}}$  & This work - DMT \\
		\hline   
	\end{tabular}
	\label{table:d6720}
	\begin{flushleft}
	$^{\textit{a}}$ This value is obtained from the catalogue of geometrical distances and presents a lower bound of 0.737~kpc and a upper bound of 0.808~kpc on the confidence interval of the estimated distance. \\
	$^{\textit{b}}$ This value is derived from the inverse {\it Gaia} parallax of 1.2707 $\pm$ 0.0594~mas. The fractional error is as low as 0.05 allowing us to derive a reliable distance using the inverse parallax. \\
	$^{\textit{c}}$ This value is derived taking into account a parallax zero point of 0.029~mas in the {\it Gaia} parallaxes using a sample of quasars \citep{lindegren2018}.\\
	$^{\textit{d}}$ This value is derived taking into account a weighted average parallax zero point of 0.051~mas in the {\it Gaia} parallaxes (see text). \\
	$^{\textit{e}}$ The error of the distance may be even higher due to the limited number of proper motion measurements.\\
	
	\end{flushleft}
\end{table}

\subsection{NGC~6302}
\label{6302a}
\begin{figure*}
	\includegraphics[width=\textwidth]{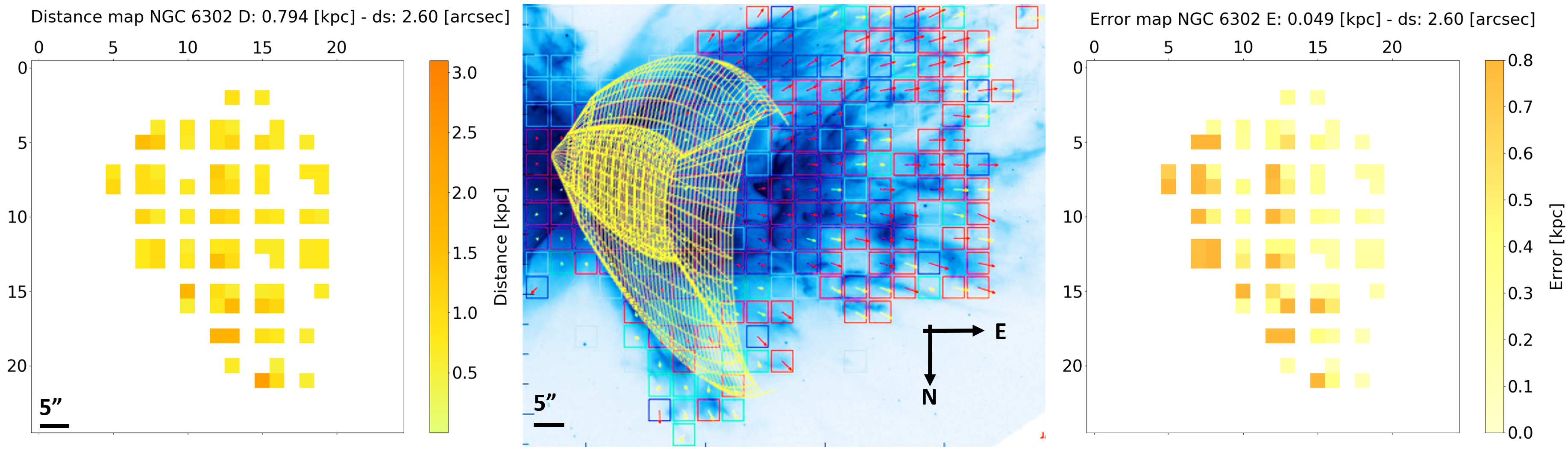}
	\caption{Distance map of NGC~6302 (left). Our model of the lobe (yellow) overlaid on its respective proper motion velocity field \citep{szyszka2011} (middle). Error map of NGC~6302 (right).}
	\label{Fig:dmt6302}
\end{figure*}

For the Butterfly Nebula there are 200 proper motion measurements available, in the [\ion{N}{ii}] emission line, with an observational error of 5~mas~yr$^{-1}$ \citep{szyszka2011}. Then, these proper motions are combined with our modelled tangential velocity field discussed in \cref{m6302} to get the distance map of NGC~6302 (\cref{Fig:dmt6302}). Besides the good number of proper motions available, the morphology of the eastern lobe is not represented very well by the DMT maps. This is the consequence of the partial coverage of the eastern lobe of the PV used to create the \textsc{shape} model (see left-hand panel of \cref{Fig:composed6302}). No apparent SDRs are identified for this nebula, only scattered cells with values above the mean distance. Perhaps a second component could explain the more extended outflow in the eastern direction, but the PV diagrams do not provide enough information to constrain this additional component.	

The distance calculated by the DMT after the application of the correction factor ($F = 1.3 \pm 0.26$, derived from the CS's surface temperature relationship; see Figure 11 in \citet[][]{schonberner2005evolution}) is 1.03~$\pm$~0.27~kpc. Similar to NGC~6720, the 1$\sigma$ distance range for NGC~6302 is not provided. Despite the few constrains in the morpho-kinematic model of NGC~6302, the distance estimation provided by the DMT is in agreement with 7 out of 12 distance estimations listed in \cref{table:d6302}. Unfortunately, no {\it Gaia} distances are available for NGC~6302. Moreover, the recent distance from \citet{FrewParker2016} is found to be almost one-third smaller than our estimation. A possible explanation for this high discrepancy is the uncertain estimation of the angular size, because of its complex bipolar morphology as well as the uncertain calculation of the integrated H$\alpha$ flux due to the contribution of the strong [\ion{N}{ii}] emission line \citep[][]{rauber2014}, which is also pointed out by \citet{FrewParker2016}.

\begin{table}
	\centering
	\caption{Derived distances to NGC~6302 in the literature.}
	\begin{tabular}{ c c }
		\hline
		Distance (kpc) & Reference \\ 
		\hline
		0.415 & \citet{maciel1980} \\ 
		2.4 & \citet{rodriguez82} \\ 
		1.7 &  \citet{rodriguez85} \\  
		1.5 &  \citet{king85} \\
		0.8 $\pm$ 0.3  & \citet{gomez89} \\
		0.91 & \citet{shaw1989} \\
		2.2 $\pm$ 1.1 & \citet{gomez89} \\
		1.6 $\pm$ 0.6 & \cite{gomez93} \\
		1.04 $\pm$ 0.16 & \citet{meaburn05} \\
		1.17 $\pm$ 0.14 & \citet{meaburn2008hubble} \\
		0.805 $\pm$ 0.14$^a$ & \cite{lago14} \\
		0.64 $\pm$0 .18 & \citet{FrewParker2016} \\
		\hline
		1.03 $\pm$ 0.27  & This work - DMT\\
		\hline   
	\end{tabular}
	\label{table:d6302}
	\begin{flushleft}
	$^a$ This distance value is not corrected for the difference between the matter and pattern velocities (correction factor F). If \textit{F} is applied, the distance should be 1.05~kpc.\\
	\end{flushleft}
\end{table}

\subsection{BD+30~3639}

Although the DMT has been previously applied to BD+30~3639 by \citet{akras2012}, we decided to apply the new version of the DMT for two reasons: (i) to compare the results between the old and new versions and (ii) to compare the results obtained from the morpho-kinematical model proposed by \citet{akras2012} and \citet{freeman2016}. The latter model is more sophisticated because it uses data from radio, infra	red, optical, X-rays, and even from molecular features such as the H$_2$ torus and the CO bullets. For this nebula, the 178  [\ion{N}{ii}] proper motion measurements from \citet{li2002angular}, with 10~percent of error, and the modelled tangential vectors from the same line of \citet{akras2012} and \citet{freeman2016} models are considered. 

\begin{figure*}
	\includegraphics[width=\textwidth]{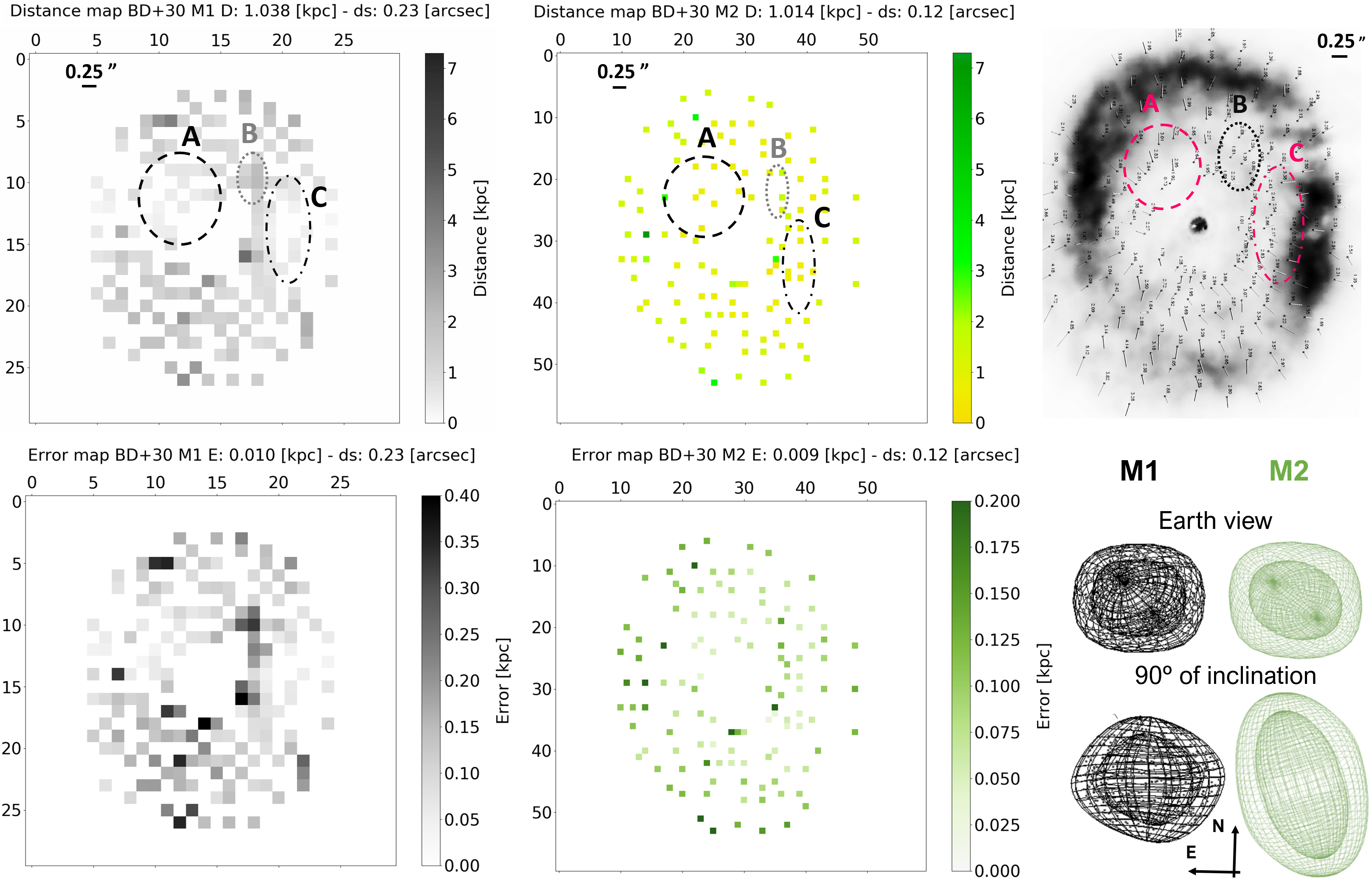}
	\caption{The distance and error maps for BD+30~3639 for the M1 model (left) and the M2 model (middle). Both models M1 and M2 share distance and error colour bars. The [\ion{N}{ii}] image of BD+30~3639 with SDRs A, B, and C overlaid (top-right). In this work, we only consider SDR A and C regions, in red color. BD+30~3639 model of \citet{akras2012} in [\ion{N}{ii}] (black) and  \citet{freeman2016} model in [\ion{N}{ii}] (green) (bottom-right).}
	\label{Fig:dmtbd+30}
\end{figure*}

There are noteworthy differences in the structure and the velocity fields between the two \textsc{shape} models. \citet{akras2012} assumed two single ellipsoidal boxy shells for the modelling of the kinematic data in the [\ion{N}{ii}] and [\ion{O}{iii}] lines (hereafter model M1, black models in the bottom-left-hand panel of \cref{Fig:dmtbd+30}), while \citet{freeman2016} modelled it as an individual elongated ellipsoid shell for [\ion{N}{ii}] and [\ion{O}{iii}] shells (hereafter model M2, green models in the bottom-left-hand panel of \cref{Fig:dmtbd+30}). Kinematically, the velocity field implemented by \citet{akras2012} is constructed with one homologous law for each shell ([\ion{N}{ii}] and [\ion{O}{iii}]) plus a cylindrical velocity component that is attributed to the high-velocity outflows. On the other hand, \citet{freeman2016} considered only one homologous law for both shells.

Despite the differences between the two models, both authors reached to a fit of the observed data and independently explain the high-velocity [\ion{O}{iii}] line, commonly associated with the bipolar outflows. The similar resultant distance values obtained with the DMT point to similar radial velocities between the two modelling approaches.

The resultant distance and error maps for M1 and M2 are presented in \cref{Fig:dmtbd+30}. The latter map is more homogeneous with smaller uncertainties. The different geometrical and kinematical approach of each
one is the reason for the differences between the model distance maps.

\citet{akras2012} identified two SDRs in their distance map (called A and B). The distances in SDR A are systematically smaller than the average distance value of the nebula, while those of SDR B are systematically larger. A possible systematic error was addressed for SDR A, because of the higher angular expansion measured in the [\ion{N}{ii}] line compared to the H$\alpha$ line. Therefore, the overestimation of the angular expansion velocities leads to an underestimation of distance. In the same way, a systematic error was also considered to be a possible explanation for SDR B, where the angular velocities are lower and the distances are overestimated. A comparison of the location of SDRs A and B with the [\ion{N}{ii}] image of BD+30~3639 does not indicate any relevant morphological features, that could explain the presence of the SDRs.

We compare the location of SDRs A and B proposed by \citet{akras2012} with our distance maps derived from the M1 and M2 models to examine for possible similarities. SDR A is present in the distance maps of both M1 and M2 models derived from the updated DTM. It encloses several cells with distances below the mean value as in the original work of \citet{akras2012}. SDR B is not noticeable in M1 and M2 maps. Additionally, another SDR is also identified in both distance maps M1 and M2, namely SDR C, which also exhibits values below the mean distance. The underestimation of the distances in SDR C is likely due to the overestimation of the angular expansion velocity in that region, similar to SDR A in \citet{akras2012}. Overall, this analysis indicates that none of the two models reproduces very well the structure and velocity field of BD+30 3639. Nevertheless, both models give distances consistent with the previous studies.

The DMT distances of BD+30~3639 are 1.35$\pm$0.22~kpc for the M1 model and 1.32$\pm$0.21~kpc for the M2 model, after applying the correction factor $F = 1.3\pm 0.2$ \citep{mellema2004expansion,schonberner2005evolution}. Both distances agree with 6 out of 18 estimations listed in \cref{table:dbd}. Only one of them is from a statistical method \citep{cahn1971}, while for the remaining five estimations, the expansion parallax method was applied. The 1$\sigma$ distance range is from 0.86 to 1.84~kpc for the M1 model and from 0.97 to 1.67~kpc for the M2 model. Both distance ranges were calculated from the dispersion of their distance maps ($\sigma_{\text{M1}} = 0.49$ and $\sigma_{\text{M2}} = 0.35$~kpc). Overall, the 1$\sigma$ distance range from each model encloses 12 and 9 out of the 21 available distances in the literature, respectively (see \cref{Fig:Tplot}).

Regardless of the structural and kinematic differences between the two models (M1 and M2), the distance estimations show a negligible difference. At a first glance, this seems to be an unexpected result, given that the two models are quite different. However, BD+30~3639 is viewed almost pole-on with an inclination angle of $20^{\circ}$ relative to the line of sight \citep{akras2012}. This implies that the more prolate ellipsoidal model of \citet{freeman2016} does not have a significant impact on the estimation of its distance as long as the velocity field of the nebula (observed PV diagrams) is well reproduced (see Fig. 8 in \citet{freeman2016}). The small differences observed in the PV diagrams from the two models have resulted in a lower dispersion for the M2 model compared to M1, but the final mean distance is in principle unaffected. Therefore, it is not possible from this analysis to decide which model better describes the structure of BD+30~3639. 

Surprisingly, the H$\alpha$ surface brightness-radius relation presented by \citet{FrewParker2016} gives a distance of 2.22~kpc, which does not agree with any of the distances published over the last 15 yr. In general, BD+30~3639 is a peculiar nebula with a Wolf-Rayet-type CS and high mass-loss rate. This is very likely responsible for the discrepancy between the DMT distances and the one derived by \citet{FrewParker2016}. Because of its peculiar CS, it is not easy to predict its evolution and obtain a precise value of the correction factor \citep{schonberner2005evolution}. This may be the reason for our lower value compared to the distances obtained from the recent \textit{Gaia} parallaxes \citep[][]{gaia2018}. In particular, the \textit{Gaia} parallax of the CS of BD+30~3639 yields a distance of 1.706~kpc, significantly higher than our DMT values. If the non-zero parallax offsets are taken into account, the corrected \textit{Gaia} distances become smaller and they better agree with our estimations (\cref{table:dbd}).

\begin{table}
	\centering
	\caption{BD+30~3639 distance determinations.}
	\begin{tabular}{ c c }
		\hline
		Distance (kpc) & Reference \\ 
		\hline
		4.15 & \citet{O'Dell1962} \\
		0.73 & \citet{daub1982} \\
		0.6 & \citet{pottasch1984} \\
		$2.8\ \pm\ ^{4.7}_{1.2}$ &  \citet{masson1989} \\  
		1.16 &  \citet{cahn1992} \\
		2.68 $\pm$ 0.81 & \citet{hajian1993} \\
		1.84 & \citet{van1994alternative} \\
		1.85 & \citet{zhang1995} \\
		1.5 $\pm$ 0.4 & \cite{kawamura1996} \\
		1.20 $\pm$ 0.12 & \citet{li2002angular} \\
		0.67 & \citet{phillips2002} \\
		2.14 & \cite{phillips2004} \\
		1.3 $\pm$ 0.2 & \cite{mellema2004expansion} \\
		2.57 & \citet{phillips2005} \\
		1.2 $\pm$ 0.2 & \citet{schonberner2005evolution} \\
		1.40$\pm$0.20/1.52$\pm$0.21$^{a}$&\citet{akras2012} \\
		2.22 $\pm$ 0.63 & \citet{FrewParker2016} \\
		1.649$^{\textit{b}}$ & \citet{Bailer2018} \\	
		1.706 $\pm$ 0.184$^{\textit{c}}$ & {This work - no zero-point correction} \\
		1.625 $\pm$ 0.175$^{\textit{d}}$ & {This work - zero-point correction A} \\
		1.569 $\pm$ 0.169$^{\textit{e}}$ & {This work - zero-point correction B}\\
		\hline
		1.35 $\pm$ 0.22 & Present work - M1 - DMT \\
		1.32 $\pm$ 0.21 & Present work - M2 - DMT\\
		0.86 - 1.84 & Most probable distance range - M1 - DMT \\
		0.97 - 1.67 & Most probable distance range - M2 - DMT \\
		\hline   
	\end{tabular}
	\label{table:dbd}
	\begin{flushleft}
	$^{a}$ These two DMT distance estimations correspond to different offsets between the geometrical centre of the nebula and the velocity field of the gas (0.25~arcsec and 0.5~arcsec, respectively).\\
	$^{\textit{b}}$This value is obtained from the catalogue of geometrical distances and it is characterized by a lower bound of 1.486~kpc and an upper bound of 1.849~kpc on the asymmetric confidence interval of the estimated distance. \\
	$^{\textit{c}}$ This value is derived from the inverse {\it Gaia} parallax of 0.5863 $\pm$ 0.0633~mas. The fractional error is as low as 0.11 allowing us to derive a reliable distance using the inverse parallax. \\
	$^{\textit{d}}$ This value is derived taking into account a parallax zero point of 0.029~mas in the {\it Gaia} parallaxes using a sample of quasars \citep{lindegren2018}.\\
	$^{\textit{e}}$ This value is derived taking into account a weighted average parallax zero point of 0.051~mas in the {\it Gaia} parallaxes (see text). \\
	
	\end{flushleft}
\end{table}

\subsection{GK~Persei}

The knotty structure of the nova remnant of GK~Persei was modelled by \citet{harvey2016} with a morpho-kinematic approach using \textsc{shape}. The model was based on high-resolution spectroscopic observations and imaging along with optical archival data from several epochs and facilities. 115 knots were modelled as individual 3D cylinders whose specific expansion velocities, distance from the CS, PA, and inclination are associated with the measured proper motions of \citet{liimets2012}. \citet{harvey2016} also modelled the overall shape of the knot distribution, not as a distorted spherical shell as \citet{liimets2012} considered, but as a cylindrical shell with its symmetry axis being related to the inclination of the central binary system. Additional bipolar features were included, to account for the knots that are not enclosed by the barrel structure. The best fit of the data indicates that the shell can be better described by a cylindrical shape, and not a spherical one \citep{harvey2016}.

\begin{figure*}
	\includegraphics[width=\textwidth]{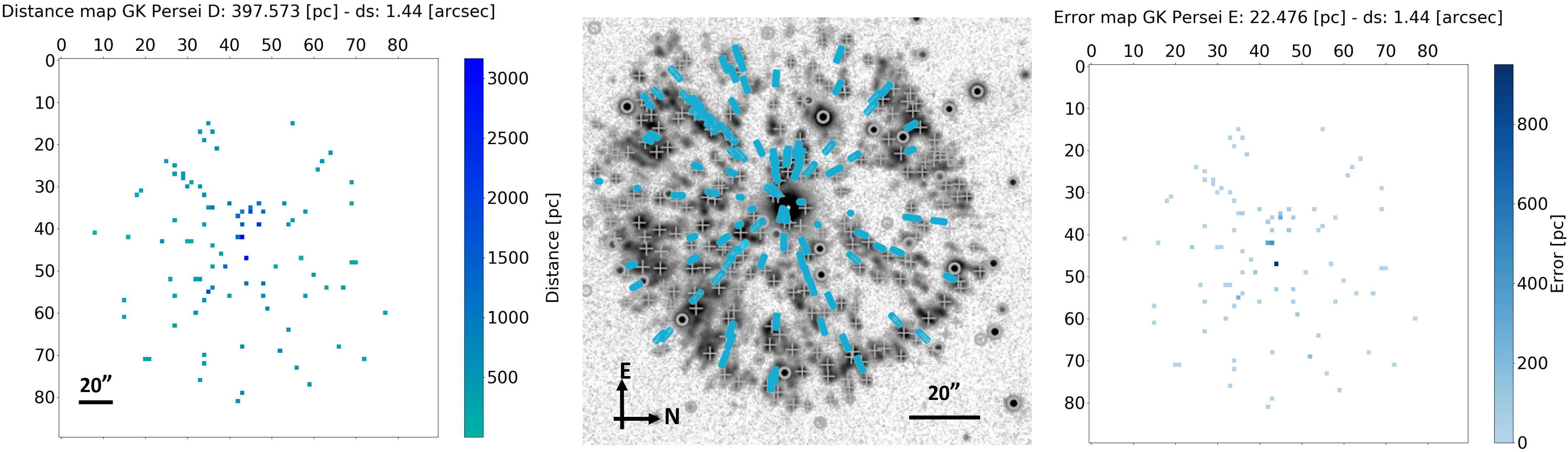}
	\caption{The distance map of GK~Persei (left-hand panel). The modelled 3D knots \citep{harvey2016} are overlaid on the H$\alpha$-[\ion{N}{ii}] image of GK~Persei \citep{liimets2012} (middle panel). The error map of GK~Persei (right-hand panel).}
	\label{Fig:dmtgk}
\end{figure*}

To apply the DMT to the remnant of GK~Persei, the 3D knotty model from \citet{harvey2016} and the 117 proper motions measured by \citet{liimets2012} in [\ion{N}{ii}] were used. In addition, observational errors for the proper motions are available, and they are taken into account in the distance determination. The axisymmetric distribution of these knots is represented in the central panel of \cref{Fig:dmtgk} overlaid on the H$\alpha$-[\ion{N}{ii}] image of GK~Persei, which displays the measured proper motions with crosses. 

The distance and error maps for GK~Persei are presented in \cref{Fig:dmtgk}, both displaying faithfully the distribution of the knots. The error map is not homogeneous and displays higher errors towards the centre of the remnant. No SDRs are identified in GK~Persei.

The resulting DMT distance for GK~Persei is 398$\pm$23~pc. Note that the correction factor (\textit{F}) has been studied only for PNe and for this reason it is not applicable to GK~Persei. In addition to the mean distance value, we also provide the 1$\sigma$ distance range for GK~Persei (203-592~pc). Our final distance for GK~Persei agrees with 5 out of the 11 distances listed in \cref{table:dGKPer}, while our 1$\sigma$ range encloses 11 of these 12 estimations (see \cref{Fig:Tplot}). Our distance is consistent with the previous expansion parallax distances of \citet{slavin1995} and \citet{downes2000}, for which no correction was applied either. Moreover, the upper limit of our distance estimation is found to be only 31~pc lower than the lower limit of the trigonometric distance of \citet{harrison2013} and 36~pc lower than the distances derived from the \textit{Gaia} parallaxes after the zero point corrections. This difference in the distance is attributed to the lack of correction factor for nova remnants, whose value in this case should be around 1.08 for a good match between our distance and those from the {\it Gaia} parallaxes \citep{gaia2018}.

\begin{table}
	\centering
	\caption{Distances calculated for GK~Persei.}
	\begin{tabular}{ c c }
		\hline
		Distance [pc) & Reference \\ 
		\hline
		470 & \citet{mclaughlin1960} \\ 
		525 & \citet{duerbeck1981} \\ 
		726 &  \citet{sherrington1983} \\  
		337 &  \citet{warner1987} \\
		455$\pm$ 30  & \citet{slavin1995} \\
		460$\ \pm\ ^{69}_{59}$ & \citet{downes2000} \\
		420 & \citet{barlow2006} \\
		400 $\pm$ 30 & \cite{liimets2012} \\
		477$\ \pm\ ^{28}_{25}$ & \citet{harrison2013} \\
	    437$^{\textit{a}}$ & \citet{Bailer2018} \\
		442 $\pm$ 8$^{\textit{b}}$ & {This work - no zero-point correction} \\
		436 $\pm$ 8$^{\textit{c}}$ & {This work - zero-point correction A} \\
		432 $\pm$ 8$^{\textit{d}}$ & {This work - zero-point correction B} \\
		\hline	
		398 $\pm$ 23  & This work - DMT \\
		203 - 592 & Most probable distance range - DMT\\
		\hline   
	\end{tabular}
	\label{table:dGKPer}
	\begin{flushleft}
	$^{\textit{a}}$ This value is obtained from the catalogue of geometrical distances and presents a lower bound of 428 pc and an upper bound of 445 pc on the confidence interval of the estimated distance.\\
	$^{\textit{b}}$ This value is derived from the inverse {\it Gaia} parallax of 2.2627 $\pm$ 0.0431~mas. The fractional error is as low as 0.02 allowing us to derive a reliable distance using the inverse parallax.\\
	$^{\textit{c}}$ This value is derived taking into account a parallax zero point of 0.029~mas in the {\it Gaia} parallaxes using a sample of quasars \citep{lindegren2018}.\\
	$^{\textit{d}}$ This value is derived taking into account a weighted average parallax zero point of 0.051~mas in the {\it Gaia} parallaxes (see text). \\
	
	\end{flushleft}
\end{table}

\section{DISCUSSION AND CONCLUSIONS}
\label{F}

In this work, we demonstrated that the distance mapping technique (DMT) is a constructive generalization of the expansion parallax method. By applying this method to PNe with several proper motion measurements available, we were able to get several distance estimations through the nebula and finally obtain a mean distance value. The first application of the DMT by \citet{akras2012} and the  nebular distance obtained in that work show the potential of DM. It can be used for getting reliable distances for expanding nebulae (PNe or even novae) if there are velocity fields available from 3D morpho-kinematic models and data sets of angular expansion velocities (proper motions).

\begin{figure*}
	\includegraphics[scale=0.3]{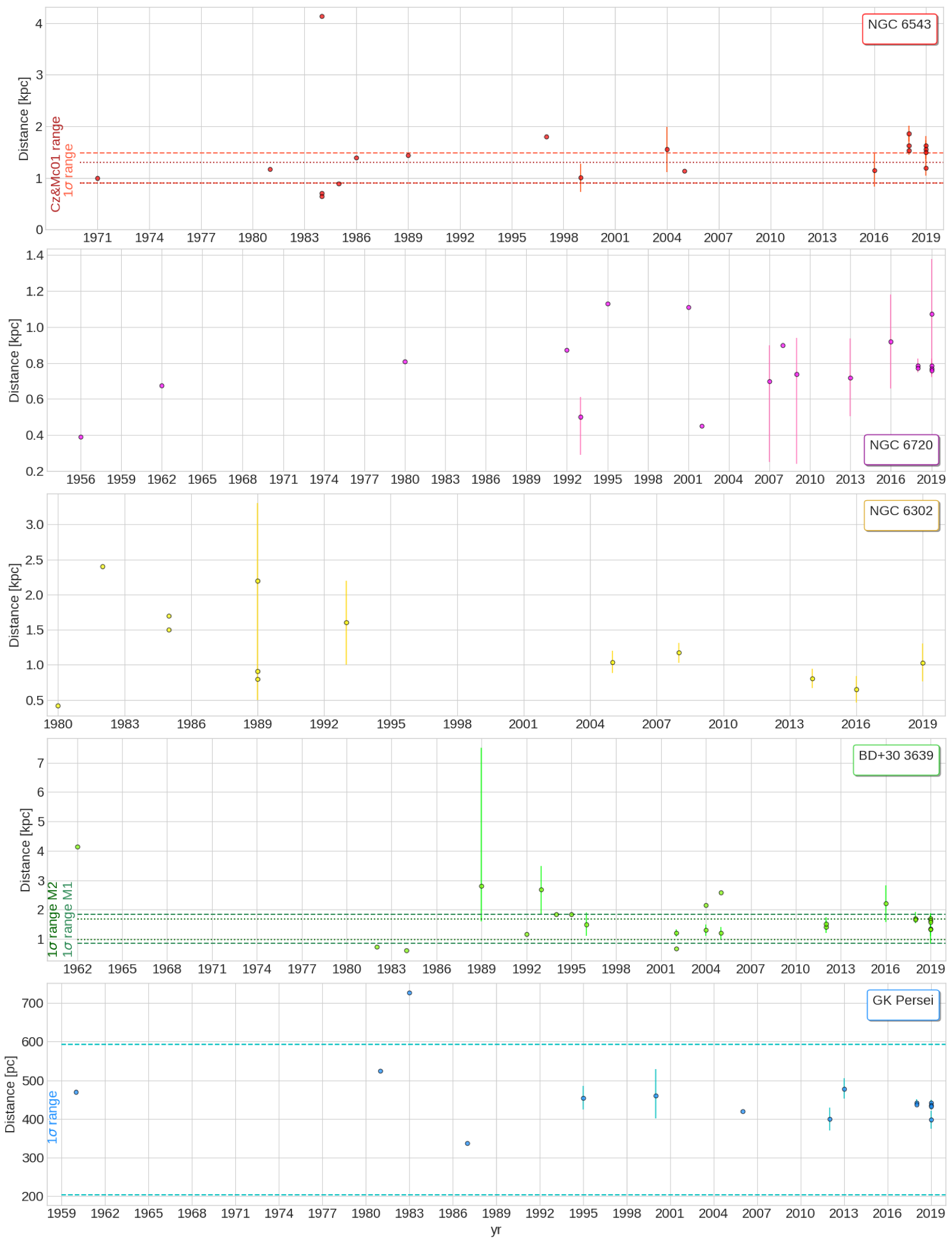}
	\caption{The distance estimations and their corresponding errors (when available), per source, through time. The dashed lines show the 1$\sigma$ distance range. The distance range of NGC~6543 derived by \citet{cazetta2001} is also provided (dotted lines). The errors of some distance estimations are smaller than the size of the markers.}
	\label{Fig:Tplot}
\end{figure*}

We applied the DMT to four PNe (NGC~6702, NGC~6543, NGC~6302 and BD+30~3639) and one nova remnant (GK Persei) and new distances were obtained for all these sources. The comparison of our DMT distances with the values from the literature shows a good agreement, as shown in \cref{Fig:Tplot}, the plot that summarizes the content of the distance tables.

By using the \textsc{shape} code, we carried out our own morpho-kinematic modelling for NGC~6543, NGC~6302, and NGC~6720. For all PNe, we consider the simplest possible structures for the purpose of this work, without specifically modelling features such as knots, jets, filaments, and structures that are generally enhanced in the low-ionization emission lines or rich in molecular gas (see \citet{goncalves2001,akrasdenise2016,akras2017}.

NGC~6543 was modelled as a bipolar structure with a homologous expansion, where the lobes are spherical shells of constant density. With the modelled velocity field and more than 100 proper motions, a distance of 1.19~$\pm$~0.15~kpc was derived using the DMT. The 1$\sigma$ distance range enclosing 50\% of the distances was previously published. Along with this, new trigonometric {\it Gaia} distances was derived in this work.

Two SDRs with distances below the mean value were identified in the distance maps of NGC~6543. These SDRs are likely results of the expansion of two different nebular components (the inner ellipse and the conjoined bubbles) leading to higher proper motions and thus lower distances. The PN NGC~6543 is the perfect nebula for a future detailed morpho-kinematic study and DMT application using the hundreds of {\it Gaia} proper motions determined from its expanding gas. These proper motions, however, were not included in the {\it Gaia} DR2 and we have to wait for the future data releases.

The model of NGC~6720 was constructed considering an elliptical prolate shell viewed pole-on. Only 22 proper motion measurements were available for this nebula. We obtained a distance of 1.07~$\pm$~0.3~kpc. Its 1$\sigma$ distance range encloses 50\% of literature values. 

A portion of the eastern lobe of the bipolar PN NGC~6302 was modelled as a conical shell. Using the modelled velocity field from our morpho-kinematic model and 200 proper motions, the resultant DMT distance was found to be equal to 1.03~$\pm$~0.27~kpc. 58\% of the previous distances agree with our value and its uncertainty.

Two morpho-kinematic models, M1 \citep{akras2012} and M2 \citep{freeman2016}, with different geometrical and kinematical approaches, were used for BD+30~3639, plus more than 100 proper motions yielding distances of 1.35 $\pm$ 0.22~kpc for M1 and 1.32 $\pm$ 0.21~kpc for M2. Both distances are in good agreement with the values in the literature. The 1$\sigma$ distance range of the M1 and M2 models encloses 60\% of the previous distance estimations. 

The performance of the new version of the DMT presented in this work has been benchmarked by comparing it with its first version via BD+30 3639. We concluded that M2 may provide a better distance representation of the nebula due to its lower dispersion and therefore a better constrained distance. The mean distances are, however, very similar.

A multicomponent \textsc{shape} model for the nova remnant GK~Persei was used in the DMT along with more than 100 proper motion measurements. A distance of 398~$\pm$~23~pc was derived together with the 1$\sigma$ range of 203-592~pc, both in good agreement with the previously published values, enclosing 92\% of the values. The wide range of distances indicates that the application of the expansion parallax method, by using only one or a few proper motion measurements in a specific region of the nebula, may not provide an accurate distance.

Four of the objects studied in this work have available \textit{Gaia} parallaxes in the DR2. The small fractional error of these parallaxes ($<$12~percent) allowed us to obtain their distances from the inverse of their parallax measurements. It was found that the \textit{Gaia} distances are higher than our DMT distances, except for NGC~6720. Nevertheless, taking into account the parallax offset in the DR2, the corrected distances become more consistent with our results.

It should also be noted that the distances obtained from the Bayesian approach \citep{Bailer2018}, using the non-corrected raw parallax measurements, were found to agree better with the distances obtained for the inverse of the corrected parallaxes than with the non-corrected ones. Until a proper study on PNe determines the parallax offset for these objects, we recommend the use of the distances from the Bayesian approach. 

A comparison between the distances derived by \cite{Schonberner2019} and the trigonometric {\it Gaia} distances shows a difference between 0.04 and 0.08 percent, which is comparable with the difference found between the corrected and non-corrected for parallax zero point {\it Gaia} distances derived in this work.

The analysis of all the distances available in the literature for the four PNe together with the new distance from the {\it Gaia} parallaxes demonstrated that different techniques result in substantially different distances. If, though, we examine the distance estimations over time, they appear to converge towards more reliable values. Yet, robust distance estimations still remain one of the most difficult tasks in the field of PNe.

\section*{Acknowledgements}
The authors thank the anonymous referee for his/her thorough review of our paper as well as his/her comments that helped us to improve the quality of the paper. The authors would also like to thank Dr. Eamonn Harvey for providing us with his \textsc{shape} model of GK~Persei and Dr. Marcus Freeman for providing us with his \textsc{shape} model of BD+30~3936. SG-G. acknowledges CAPES, the Brazilian Federal Agency for Support and Evaluation of Graduate Education within the Ministry of Education of Brazil. SA and DRG thank support of the CNPq, Conselho Nacional de Desenvolvimento Cient\'{i}fico e Tecnol\'{o}gico, respectively from grants 300336/2016-0 and 304184/2016-0. WS acknowledges support from DGAPA (Direcci\'on General de Asuntos del Personal Ac\'ademico), UNAM grant PAPIIT (Programa de Apoyo a Proyectos de Investigaci\'on e Innovaci\'on Tecnol\'ogica) 104017. Based on observations made with the NASA/ESA Hubble Space Telescope, obtained from the data archive at the Space Telescope Science Institute. STScI is operated by the Association of Universities for Research in Astronomy, Inc. under NASA contract NAS 5-26555. This work presents results from the European Space Agency (ESA) space mission Gaia. Gaia data are being processed by the Gaia Data Processing and Analysis Consortium (DPAC). Funding for the DPAC is provided by national institutions, in particular the institutions participating in the Gaia MultiLateral Agreement (MLA). The Gaia mission website is https://www.cosmos.esa.int/Gaia. The Gaia archive website is https://archives.esac.esa.int/Gaia. Finally, this publication makes use of many software packages in Python, including: \textsc{matplotlib} \citep{Hunter2007}, \textsc{numpy} \citep{Walt2011} and \textsc{scipy} \citep{jones2001}.
	



\bibliographystyle{mnras}
\bibliography{references} 




%
%


\bsp	
\label{lastpage}
\end{document}